\documentclass{article}

\usepackage{spconf,amsmath,amssymb,graphicx,booktabs,multirow,array,bm,url,hyperref,etoolbox}
\apptocmd{\thebibliography}{\setlength{\itemsep}{0pt}\setlength{\parsep}{0pt}}{}{}

\graphicspath{{figures/}}
\newcommand{\method}{PrecipJEPA}
\newcommand{\hjepa}{H-JEPA}
\newcommand{\tfp}{TFP}
\newcommand{\pmsr}{PMSR}

\newif\ifincludereferences
\includereferencestrue

\title{PrecipJEPA: JEPA-Regularized Future-State Prediction with Motion--Source Rendering for Precipitation Nowcasting}
\name{\fontsize{11}{13}\selectfont Yufeng Zhu$^{1}$, Dan Niu$^{1}$, Qiliang Wu$^{2}$,
Weiwei Huang$^{2}$, Yixiao Liang$^{2}$, Yongchao Feng$^{3}$,
Chunlei Shi$^{1,*}$}
\address{\fontsize{9}{10.5}\selectfont
$^{1}$Department of Automation, Southeast University, Nanjing 210096, China;
$^{*}$Corresponding author: Chunlei Shi \\
\fontsize{9}{10.5}\selectfont $^{2}$Beijing Fengyun Meteorological Science and Technology Development Co., Ltd., Beijing 100081, China \\
\fontsize{9}{10.5}\selectfont $^{3}$State Key Laboratory of Virtual Reality Technology and Systems, Beihang University, Beijing 100191, China}

\begin{document}
\ninept

\maketitle

\begin{abstract}
Long-term precipitation nowcasting requires modeling radar-echo evolution
while preserving localized high-intensity structures. Recent radar-specific
studies motivate location-aware prediction and separating echo displacement
from intensity change. However existing encoders learn historical
representations mainly from final forecast errors. We propose PrecipJEPA, which
couples a structured forecasting path with an auxiliary path that enriches its
encoder from observed radar history. In the forecasting path, an online encoder
first converts the observations into spatiotemporal tokens. The Task-Driven Future-State Predictor
(TFP) combines these tokens with a recent-dynamics summary and spatiotemporal
queries to construct future radar states. The Parallel Motion--Source Renderer
(PMSR) decodes these states into motion and source--sink fields that transform
the latest observation into future frames. During joint training, the
History-Masked JEPA (H-JEPA) operates on the auxiliary path to predict masked
historical features from visible context, directly supervising the same online
encoder from the observed sequence. Experiments on
SEVIR and MeteoNet show that PrecipJEPA improves highest-threshold CSI by
118.6\% and 35.1\%, respectively, over the strongest baselines, while
maintaining the highest mean CSI throughout the 3-hour forecast.

\end{abstract}


% ICASSP-style compact body. Prior-work positioning is integrated into the
% Introduction and Method rather than placed in a standalone Related Work.
\section{Introduction}
\label{sec:introduction}

Precipitation nowcasting predicts future radar echoes from recent observations
and provides short-term guidance for severe weather warnings \cite{NIPS2015_07563a3f,LI2026130775,10.3389/feart.2022.846113}.  
As lead time increases, echo motion, deformation, intensification, and dissipation become
harder to predict. Reliable long-term nowcasting must summarize
observed echo evolution and translate it into coherent future radar fields.

To model this evolution, deep-learning methods for precipitation nowcasting
have been developed along deterministic and generative directions
\cite{LI2026130775}.  Early
deterministic models combined convolutional feature extraction with recurrent
memory to capture radar evolution
\cite{NIPS2015_07563a3f,NIPS2017_a6db4ed0,
wang2022predrnnrecurrentneuralnetwork}.  Later
convolutional and Transformer architectures expanded the spatial context and
modeled longer-range dependencies
\cite{Gao_2022_CVPR,NEURIPS2022_a2affd71}.  Meanwhile, generative and
reconstruction-oriented methods use adversarial learning
\cite{ravuri2021skilful,zhang2023skilful}, diffusion models
\cite{NEURIPS2023_f82ba6a6,Yu_2024_CVPR,
gong2024cascastskillfulhighresolutionprecipitation,shi2026wavec2r}, and flow
matching \cite{ribeiro2026flowcastadvancingprecipitationnowcasting} to
represent forecast variability and recover fine-scale structure.  Building on
these advances, recent
radar-specific models further adapt their forecasting architectures to the
properties of radar echoes.

AlphaPre separates amplitude and phase to model changes in precipitation
intensity and position \cite{lin2025alphapre}, whereas exPreCast focuses on
local precipitation patterns to preserve fine-scale radar structure
\cite{ICLR2026_b16bff5d}.  Taken together, these studies motivate a
causal-inference-inspired decomposition of radar evolution that preserves local
spatial structure while distinguishing echo displacement from intensity change.  Meeting this
requirement over a long forecast also calls for future states that remain
specific to time and location.

Such a forecasting design, however, can only use information retained by its
encoder.  If the encoder is optimized solely through the final forecasting
loss, its historical features are learned indirectly from output errors, with
no explicit objective for predictive relationships within the observed
sequence.  This motivates an additional learning objective for the history
itself.  The Joint-Embedding Predictive Architecture (JEPA), developed along
the broader direction of latent world modeling, offers such an objective.
I-JEPA \cite{10205476} and V-JEPA
\cite{bardes2024revisitingfeaturepredictionlearning} predict masked target
representations from visible context in embedding space rather than
reconstructing every pixel.  Since the target of JEPA
is produced by a learned encoder, the prediction can focus on shared structure
instead of reproducing every local pixel variation.  Applied to radar history,
this principle trains the encoder to infer masked echo representations from
spatiotemporal context, complementing forecast supervision.

Bringing these ideas together, we propose \method{}, which couples a structured
forecasting path with a history-based auxiliary path.  The forecasting path
encodes the observed sequence into spatiotemporal tokens.  The Task-Driven
Future-State Predictor (\tfp{}) combines this representation with a
recent-dynamics summary, guided by separate temporal and spatial queries, to
construct location-specific states at every lead time.  The Parallel
Motion--Source Renderer (\pmsr{}) decodes the states
into motion and source--sink fields that transport the latest observation and
model local echo growth and decay.  During training, the History-Masked JEPA
(\hjepa{}) forms the auxiliary path and predicts masked latent targets from
visible history.  Combined with the forecasting loss, its objective guides the
online encoder using both observed evolution and future radar targets.  From
this representation, \tfp{} and \pmsr{} generate coherent radar forecasts that
better preserve high-intensity echoes at later lead times.

\begin{figure*}[t]
    \centering
    \includegraphics[width=\textwidth]{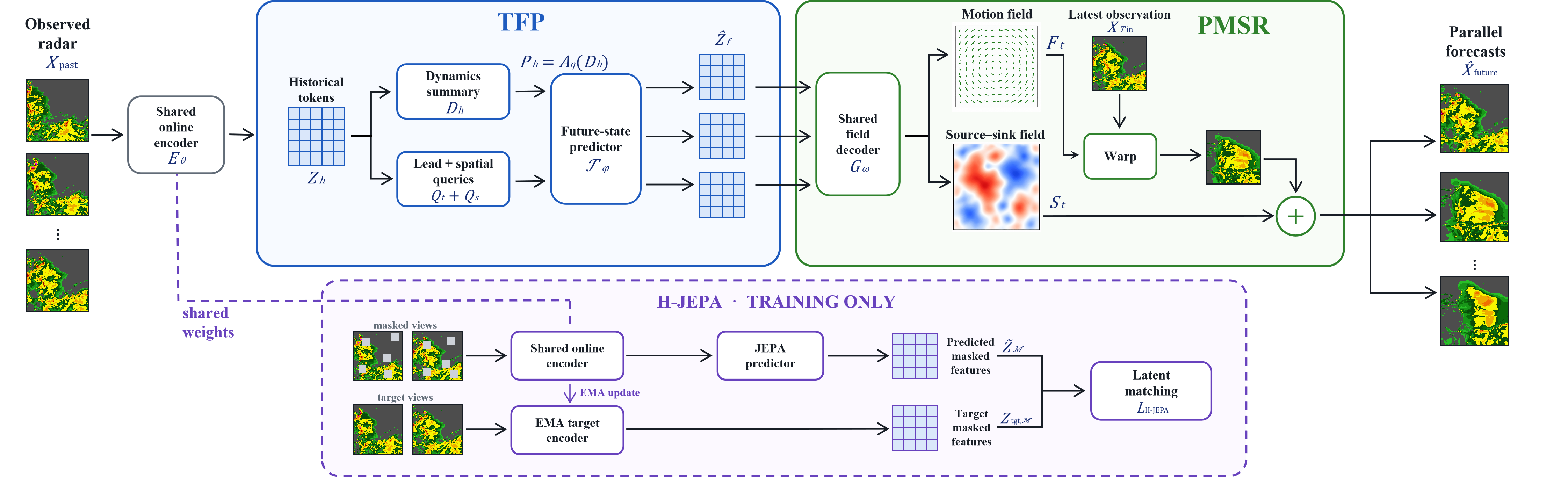}
    \caption{Overview of \method{}.  \tfp{} maps historical representations
    to lead-specific future states, which \pmsr{} renders into radar
    forecasts, while \hjepa{} supervises the shared encoder through
    masked-latent prediction during training.}
    \label{fig:precipjepa_framework}
\end{figure*}

Our contributions are summarized below.
\begin{itemize}
    \setlength{\itemsep}{1pt}
    \setlength{\parskip}{0pt}
    \setlength{\parsep}{0pt}
    \setlength{\topsep}{2pt}
    \item We propose \method{}, which combines a structured forecasting path
    with a history-masked learning path for coherent long-term forecasts and
    improved high-threshold skill.
    \item We design \tfp{} and \pmsr{} for future-state construction and
    motion--source rendering, while joint \hjepa{} training strengthens the
    encoder's modeling of echo evolution.
    \item Experiments on SEVIR \cite{NEURIPS2020_fa78a161} and MeteoNet
    \cite{larvor2020meteonet} improve CSI at the highest thresholds by 118.6\%
    and 35.1\% over the strongest baselines; ablations validate the design.
\end{itemize}

% TODO: Revise the third contribution with verified public-baseline,
% repeated-seed, cross-dataset, and efficiency evidence when those experiments
% are complete.

\section{Method}
\label{sec:method}

\subsection{Problem Formulation and Overview}
\label{sec:problem_formulation}

Given an observed radar sequence
$X_{\mathrm{past}}=\{X_t\}_{t=1}^{T_{\mathrm{in}}}$, precipitation
nowcasting aims to predict
$X_{\mathrm{future}}=\{X_{T_{\mathrm{in}}+t}\}_{t=1}^{T_{\mathrm{out}}}$,
where $X_t\in[0,1]^{H\times W\times C}$.  The model learns the mapping
\begin{equation}
    \hat X_{\mathrm{future}}
    =\mathcal{P}(X_{\mathrm{past}}),
    \label{eq:forecast_mapping}
\end{equation}
where $\hat X_{\mathrm{future}}$ is the predicted radar sequence.

Figure~\ref{fig:precipjepa_framework} summarizes \method{}, which contains a
main forecasting path and a history-only auxiliary branch.  In the main path,
an online encoder $E_{\theta}$ represents the
observed sequence as spatiotemporal tokens $Z_h$.  \tfp{} converts this
historical representation into future states $\hat Z_f$, and \pmsr{} renders
the corresponding radar sequence from these states.
To provide an additional learning signal for the historical representation,
\hjepa{} predicts masked features within the observed sequence during training
and shares the online encoder with the main path.

% Current values are single-run results. Replace them with repeated-seed
% statistics if those runs are completed before submission.
\begin{table*}[t]
  \centering
  \caption{Quantitative comparison on SEVIR and MeteoNet. Aggregate metrics
  cover all thresholds and forecast times; CSI-last covers the final forecast
  hour. CSI$_{181}$, CSI$_{219}$, CSI$_{24}$, and CSI$_{32}$ report the two
  highest standard thresholds of each dataset.
  Best results are in \textbf{bold} and second-best results are
  \underline{underlined}.}
  \label{tab:main_comparison}
  \setlength{\tabcolsep}{0.8pt}
  \renewcommand{\arraystretch}{0.85}
  \resizebox{\textwidth}{!}{%
  \begin{tabular}{@{}l|cccccccc|cccccccc@{}}
    \toprule
    & \multicolumn{8}{c|}{\textbf{SEVIR}}
    & \multicolumn{8}{c}{\textbf{MeteoNet}} \\
    \cmidrule(lr){2-9}\cmidrule(lr){10-17}
    Method
    & CSI$\uparrow$ & CSI-p4$\uparrow$ & CSI-p16$\uparrow$ & HSS$\uparrow$
    & LPIPS$\downarrow$ & CSI-last$\uparrow$
    & CSI$_{181}$$\uparrow$ & CSI$_{219}$$\uparrow$
    & CSI$\uparrow$ & CSI-p4$\uparrow$ & CSI-p16$\uparrow$ & HSS$\uparrow$
    & LPIPS$\downarrow$ & CSI-last$\uparrow$
    & CSI$_{24}$$\uparrow$ & CSI$_{32}$$\uparrow$ \\
    \midrule
    SimVP~\cite{Gao_2022_CVPR}
    & 0.2073 & 0.2277 & 0.2405 & 0.2518 & 0.4348 & 0.1362 & 0.0276 & 0.0147
    & 0.2598 & 0.3251 & 0.3562 & 0.3589 & 0.3444 & 0.1290 & 0.1974 & 0.0674 \\
    Earthformer~\cite{NEURIPS2022_a2affd71}
    & 0.2006 & 0.2193 & 0.2384 & 0.2456 & 0.4434 & 0.1366 & 0.0168 & 0.0024
    & 0.2398 & 0.2968 & 0.3393 & 0.3370 & 0.4130 & 0.1309 & 0.1924 & 0.0362 \\
    AlphaPre~\cite{lin2025alphapre}
    & \underline{0.2162} & \underline{0.2381} & 0.2495 & 0.2652 & 0.4192 & 0.1432 & 0.0311 & 0.0139
    & \underline{0.2708} & \underline{0.3322} & 0.3624 & \underline{0.3723} & 0.3308 & 0.1366 & 0.2191 & 0.0771 \\
    exPreCast~\cite{ICLR2026_b16bff5d}
    & 0.2142 & 0.2365 & \underline{0.2497} & \underline{0.2726} & 0.3617 & \underline{0.1472} & \underline{0.0391} & 0.0161
    & 0.2583 & 0.3230 & \underline{0.3668} & 0.3659 & 0.3170 & 0.1532 & \underline{0.2279} & \underline{0.1043} \\
    CasCast~\cite{gong2024cascastskillfulhighresolutionprecipitation}
    & 0.1977 & 0.2105 & 0.2333 & 0.2549 & 0.4108 & 0.1341 & 0.0337 & 0.0145
    & 0.2191 & 0.2299 & 0.2224 & 0.3131 & 0.3454 & \underline{0.1596} & 0.2053 & 0.0824 \\
    DiffCast~\cite{Yu_2024_CVPR}
    & 0.2036 & 0.2191 & 0.2299 & 0.2620 & \underline{0.3592} & 0.1399 & 0.0381 & \underline{0.0172}
    & 0.2511 & 0.3076 & 0.3379 & 0.3572 & \underline{0.2926} & 0.1580 & 0.2268 & 0.1024 \\
    \midrule
    \textbf{PrecipJEPA}
    & \textbf{0.2315} & \textbf{0.2512} & \textbf{0.2510} & \textbf{0.2974}
    & \textbf{0.3347} & \textbf{0.1493}
    & \textbf{0.0591} & \textbf{0.0376}
    & \textbf{0.2887} & \textbf{0.3470} & \textbf{0.3817} & \textbf{0.4043}
    & \textbf{0.2317} & \textbf{0.1734}
    & \textbf{0.2694} & \textbf{0.1409} \\
    \bottomrule
  \end{tabular}%
  }
\end{table*}

\begin{figure*}[t]
    \centering
    \includegraphics[width=\textwidth]{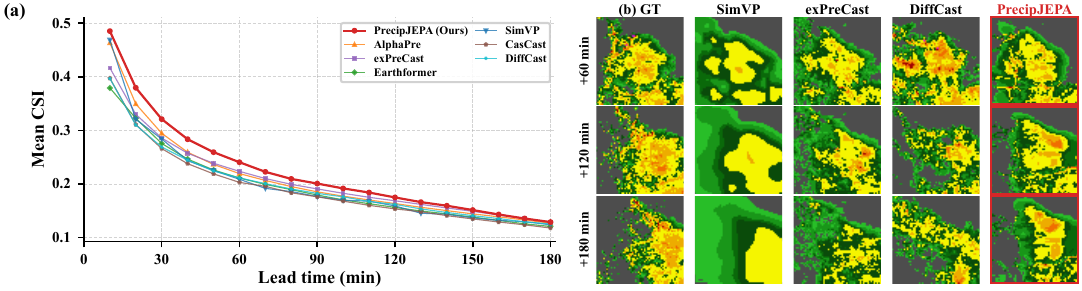}
    \caption{Long-term evaluation on SEVIR: (a) mean CSI over the full
    180-minute forecast; (b) example forecasts from different models.}
    \label{fig:sevir_long_horizon}
\end{figure*}

\subsection{Task-Driven Future-State Predictor (\tfp{})}
\label{sec:future_states}

Predicting future radar fields requires the model to describe how
the observed evolution may unfold at different lead times and spatial
locations.  \tfp{} therefore constructs a separate latent state for each
future time--location pair.  As shown in
Fig.~\ref{fig:precipjepa_framework}, the online encoder maps the observed
sequence $X_{\mathrm{past}}$ to historical tokens
$Z_h=E_{\theta}(X_{\mathrm{past}})$.  \tfp{} processes $Z_h$ through two
complementary paths.  The dynamics path derives an observation-dependent
prompt, while the query path represents each future time and spatial location.

The dynamics path summarizes the evolution recorded in the historical tokens.
With $Z_{h,t}$ denoting the tokens at time $t$, the dynamics summary is
\begin{equation}
\begin{aligned}
    \Delta Z_h&=Z_{h,T_{\mathrm{in}}}-Z_{h,T_{\mathrm{in}}-1},\\
    \bar A_h&=\frac{1}{T_{\mathrm{in}}-1}
    \sum_{t=2}^{T_{\mathrm{in}}}|Z_{h,t}-Z_{h,t-1}|,\\
    D_h&=[Z_{h,T_{\mathrm{in}}},\;\Delta Z_h,\;\bar A_h].
\end{aligned}
\label{eq:dynamics_summary}
\end{equation}
These terms describe the latest historical state, its most recent change, and
the average activity over the observed period.  A prompt encoder transforms
this summary into $P_h=A_{\eta}(D_h)$, which provides
observation-specific evolution cues.

The query path assigns a learned query to each future time and spatial location.
The temporal query $Q^{\mathrm{t}}$ specifies the forecast time, whereas the
spatial query $Q^{\mathrm{s}}$ specifies the corresponding radar region.
Finally, the predictor $\mathcal{T}_{\phi}$ combines their sum with the historical
tokens and dynamics prompt,
\begin{equation}
    \hat Z_f
    =\mathcal{T}_{\phi}
    (Q^{\mathrm{t}}+Q^{\mathrm{s}};Z_h,P_h).
    \label{eq:future_state_prediction}
\end{equation}
The resulting future states are subsequently rendered into radar frames by
\pmsr{}.

\subsection{Parallel Motion--Source Rendering (\pmsr{})}
\label{sec:motion_source}

The future states $\hat Z_f$ describe the expected radar evolution in latent
space and must be converted into observable radar fields.  A shared decoder
$G_{\omega}$ separates these states into raw motion and source--sink fields
$\tilde F$ and $\tilde S$.  Rather than decoding future intensities directly,
\pmsr{} uses these fields to form two rendering paths.  At lead time $t$, the
transport path uses the motion field $F_t$ to displace echoes in the latest
observation, while the source path uses $S_t$ to describe local intensity
changes after transport.  This decomposition represents echo displacement and
intensity variation with separate variables while retaining the latest
observed field as the rendering reference.  The decoded fields are bounded as
\begin{equation}
\begin{aligned}
    (\tilde F,\tilde S)&=G_{\omega}(\hat Z_f),\\
    (F,S)&=\big(F_{\max}\tanh(\tilde F),\,
    S_{\max}\tanh(\tilde S)\big).
\end{aligned}
    \label{eq:motion_source_fields}
\end{equation}
where $F_{\max}$ and $S_{\max}$ set the maximum displacement and intensity
update.

For each lead time, the transport path applies differentiable bilinear warping
to the latest observed frame $X_{T_{\mathrm{in}}}$ using $F_t$
\cite{gmd-12-1387-2019,pulkkinen2019pysteps}, while the source path adds $S_t$
to the transported result.  Their combination produces the predicted radar
frame,
\begin{equation}
    \hat X_{T_{\mathrm{in}}+t}
    =\Pi_{[0,1]}\!\left[
    \mathcal{W}(X_{T_{\mathrm{in}}},F_t)+S_t\right],
    \quad t=1,\ldots,T_{\mathrm{out}},
    \label{eq:motion_source_rendering}
\end{equation}
where $\mathcal{W}$ denotes differentiable bilinear warping and $\Pi_{[0,1]}$
clips the prediction to the valid radar range.

Through this motion--source decomposition, \pmsr{} preserves the transported
echo structure through warping, while $S_t$ accounts for local growth or decay
that displacement alone cannot represent.  This design allows the resulting
forecasts to capture changes in echo position and intensity within a unified
rendering process, thereby improving forecast skill for high-intensity
precipitation.

\subsection{Historical Representation Learning with \hjepa{}}
\label{sec:history_jepa}

Inspired by latent feature prediction in the JEPA family
\cite{10205476,bardes2024revisitingfeaturepredictionlearning}, \hjepa{} is an
auxiliary training branch that improves the historical representation used by
the forecasting path.  It creates two views of the same observed sequence.
The online encoder receives a partially masked history and extracts the
visible context.  An exponential-moving-average (EMA) target encoder observes
the complete history and supplies stable representations at the masked
positions.  A predictor then infers these target representations from the
visible context.  Since the online encoder is shared with the forecasting
path, this objective directly shapes the historical state passed to \tfp{}.

During training, we hide several spatiotemporal blocks from the observed
history.  The shared online encoder reads the remaining patches, and the
predictor estimates the latent representations of the hidden patches.  The
corresponding targets are produced by the EMA encoder from the complete
observed history.  If $\mathcal{M}$ denotes the hidden positions,
$\tilde Z_{\mathcal M}$ the predicted representations, and
$Z_{\mathrm{tgt},\mathcal M}$ their targets, the auxiliary loss is
\begin{equation}
    \mathcal{L}_{\mathrm{H\text{-}JEPA}}
    =\operatorname{MSE}\!\left(
    \operatorname{LN}(\tilde Z_{\mathcal M}),
    \operatorname{LN}(Z_{\mathrm{tgt},\mathcal M})\right).
    \label{eq:history_jepa_loss}
\end{equation}
The target encoder is updated from the online encoder by
$\bar\theta\leftarrow\mu\bar\theta+(1-\mu)\theta$ rather than by
backpropagation, where $\mu$ is the EMA decay.

\subsection{Joint Objective and Inference}
\label{sec:objective_inference}

After defining the forecasting path and the \hjepa{} branch, we optimize them
through a joint objective.  The forecasting path uses the shared online encoder
on the complete, unmasked history to obtain
$Z_h=E_{\theta}(X_{\mathrm{past}})$.  Consequently, the forecasting objective
and \hjepa{} jointly train the encoder that supplies this representation to
\tfp{}.  Joint optimization lets $Z_h$ learn from both masked historical
prediction and future forecast supervision, reducing the risk that a
subsequent forecast-only stage weakens the predictive relationships learned
from history.

The future radar sequence produced by the main path is supervised with mean
squared error
\begin{equation}
    \mathcal{L}_{\mathrm{fore}}
    =\operatorname{MSE}\!\left(
    \hat X_{\mathrm{future}},X_{\mathrm{future}}\right),
    \label{eq:forecast_objective}
\end{equation}
which averages the squared prediction error over all forecast times and spatial
locations.  The full training objective combines this forecast loss with the
historical latent loss,
\begin{equation}
    \mathcal{L}
    =\mathcal{L}_{\mathrm{fore}}
    +\lambda_J\mathcal{L}_{\mathrm{H\text{-}JEPA}},
    \label{eq:total_objective}
\end{equation}
where $\lambda_J$ balances the auxiliary objective.  At inference, the online
encoder processes $X_{\mathrm{past}}$, \tfp{} constructs $\hat Z_f$, and
\pmsr{} renders $\hat X_{\mathrm{future}}$, while the mask, \hjepa{} predictor,
and target encoder are omitted.

% Current values are single-run results. Replace them with repeated-seed
% statistics if those runs are completed before submission.
\begin{table}[t]
  \centering
  \caption{Component ablation on SEVIR and MeteoNet. H-J denotes H-JEPA. In M0 and M1,
  TFP is replaced by the basic future-state predictor; a dash under PMSR
  indicates use of the residual head. Best results are in \textbf{bold}.}
  \label{tab:factorial_ablation}
  \textbf{(a) SEVIR}\par\vspace{1pt}
  \setlength{\tabcolsep}{0.7pt}
  \resizebox{\columnwidth}{!}{%
    \begin{tabular}{@{}lccc|rrrrr@{}}
      \toprule
      Variant & TFP & PMSR & H-J & CSI$\uparrow$ & HSS$\uparrow$
      & CSI-last$\uparrow$
      & CSI$_{219}$$\uparrow$ & HSS$_{219}$$\uparrow$ \\
      \midrule
      M0 (Base) & -- & -- & -- & 0.2082 & 0.2658
      & 0.1477 & 0.0130 & 0.0239 \\
      M1 (w/o TFP) & -- & \checkmark & \checkmark & 0.2162 & 0.2761
      & 0.1344 & 0.0305 & 0.0528 \\
      M2 (w/o PMSR) & \checkmark & -- & \checkmark & 0.2082 & 0.2662
      & 0.1441 & 0.0135 & 0.0247 \\
      M3 (w/o H-J) & \checkmark & \checkmark & -- & 0.2215 & 0.2841
      & 0.1374 & 0.0342 & 0.0592 \\
      \textbf{PrecipJEPA} & \checkmark & \checkmark & \checkmark
      & \textbf{0.2315} & \textbf{0.2974}
      & \textbf{0.1493} & \textbf{0.0376} & \textbf{0.0657} \\
      \bottomrule
    \end{tabular}%
  }

  \vspace{4pt}
  \textbf{(b) MeteoNet}\par\vspace{1pt}
  \resizebox{\columnwidth}{!}{%
    \begin{tabular}{@{}lccc|rrrrr@{}}
      \toprule
      Variant & TFP & PMSR & H-J & CSI$\uparrow$ & HSS$\uparrow$
      & CSI-last$\uparrow$
      & CSI$_{32}$$\uparrow$ & HSS$_{32}$$\uparrow$ \\
      \midrule
      M0 (Base) & -- & -- & -- & 0.2473 & 0.3555
      & 0.1368 & 0.1068 & 0.1772 \\
      M1 (w/o TFP) & -- & \checkmark & \checkmark & 0.2665 & 0.3756
      & 0.1343 & 0.1246 & 0.2012 \\
      M2 (w/o PMSR) & \checkmark & -- & \checkmark & 0.2531 & 0.3634
      & 0.1438 & 0.1126 & 0.1868 \\
      M3 (w/o H-J) & \checkmark & \checkmark & -- & 0.2722 & 0.3829
      & 0.1402 & 0.1286 & 0.2067 \\
      \textbf{PrecipJEPA} & \checkmark & \checkmark & \checkmark
      & \textbf{0.2887} & \textbf{0.4043}
      & \textbf{0.1734} & \textbf{0.1409} & \textbf{0.2268} \\
      \bottomrule
    \end{tabular}%
  }
\end{table}

\begin{figure}[t]
    \centering
    \includegraphics[width=\columnwidth]{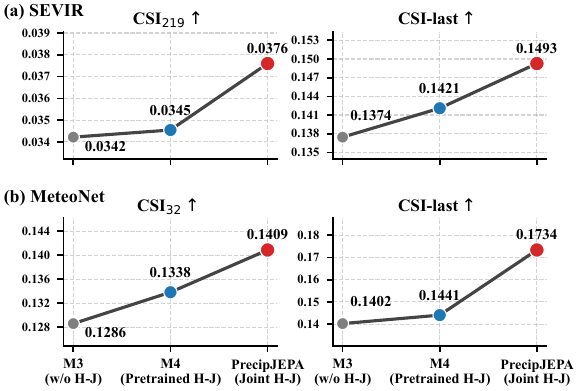}
    \caption{H-JEPA training strategies on (a) SEVIR and (b) MeteoNet. M3
    trains without H-JEPA, M4 uses it for encoder pretraining before
    forecast-only training, and \method{} jointly trains H-JEPA with the
    forecasting path.}
    \label{fig:jepa_training_strategy}
\end{figure}

\section{Experiments}
\label{sec:experiments}

\subsection{Experimental Setup}
\label{sec:experimental_protocol}

\noindent\textbf{Datasets and protocol.}
We evaluate \method{} on SEVIR~\cite{NEURIPS2020_fa78a161} and
MeteoNet~\cite{larvor2020meteonet}.  Both datasets are split chronologically.
Under a unified protocol, all models use 12 frames to forecast the following
36 frames at 5-min intervals.  Following DiffCast~\cite{Yu_2024_CVPR}, we
resize every frame to $128\times128$ to reduce GPU memory consumption.

\noindent\textbf{Metrics.}
Following DiffCast~\cite{Yu_2024_CVPR}, evaluation uses the Critical Success
Index (CSI) and Heidke Skill Score (HSS) at thresholds \{16, 74, 133, 160,
181, 219\} for SEVIR and \{12, 18, 24, 32\} for MeteoNet, together with
CSI-p4, CSI-p16, and Learned Perceptual Image Patch Similarity (LPIPS).

\noindent\textbf{Implementation and baselines.}
We train \method{} and retrain every baseline under the same input-output
protocol, spatial resolution, data split, and evaluation code, selecting
checkpoints by validation CSI. Baselines include four deterministic predictors,
SimVP~\cite{Gao_2022_CVPR},
Earthformer~\cite{NEURIPS2022_a2affd71}, AlphaPre~\cite{lin2025alphapre}, and
exPreCast~\cite{ICLR2026_b16bff5d}, as well as the two-stage generative models
CasCast~\cite{gong2024cascastskillfulhighresolutionprecipitation} and
DiffCast~\cite{Yu_2024_CVPR}. CasCast and DiffCast use SimVP as their first-stage
predictor.

\subsection{Comparison with State-of-the-Art Methods}
\label{sec:overall_results}

Table~\ref{tab:main_comparison} compares \method{} with baseline methods on
the SEVIR and MeteoNet datasets.  \method{} obtains the best value for every
reported metric on both datasets.  \method{} improves overall CSI over
AlphaPre by 7.1\% on SEVIR and 6.6\% on MeteoNet, while reducing LPIPS
relative to DiffCast by 6.8\% and 20.8\%, respectively.  The advantage is
especially clear for high-intensity precipitation: at the highest evaluation threshold,
\method{} surpasses the strongest baseline in CSI by 118.6\% on SEVIR and
35.1\% on MeteoNet.

In terms of long-term forecasting, \method{} achieves the highest CSI-last,
reaching 0.1493 on SEVIR and 0.1734 on MeteoNet, as reported in
Table~\ref{tab:main_comparison}.  The lead-time curves and visual forecasts in
Fig.~\ref{fig:sevir_long_horizon} further illustrate its performance throughout
the full 180-minute forecast.  \method{} maintains the highest mean CSI across
all evaluated lead times, while retaining more complete high-intensity echo structures in
the later forecasts.

\subsection{Ablation Study}
\label{sec:ablation}

Table~\ref{tab:factorial_ablation} examines how the three components affect
forecast behavior across both datasets. Replacing TFP with an MLP predictor in
M1 primarily hurts final-hour accuracy. Reintroducing TFP raises CSI-last from
0.1344 to 0.1493 on SEVIR and from 0.1343 to 0.1734 on MeteoNet. This consistent
recovery indicates that lead- and location-specific future states provide a
stronger basis for long-horizon rendering. The clearest contribution of PMSR
appears at high precipitation intensities. Compared with the residual head in
M2, \method{} improves CSI$_{219}$ by 178.5\% and HSS$_{219}$ by 166.0\% on
SEVIR. On MeteoNet, CSI$_{32}$ rises from 0.1126 to 0.1409, while HSS$_{32}$
increases from 0.1868 to 0.2268. Separating echo displacement from local
intensity evolution is therefore particularly valuable for retaining intense
precipitation. H-JEPA adds a complementary benefit. Removing it in M3 lowers
CSI-last from 0.1493 to 0.1374 on SEVIR and from 0.1734 to 0.1402 on MeteoNet.
The decline on both datasets shows that TFP and PMSR benefit from a shared
encoder that also learns predictive relationships within the observed history.

\begin{samepage}
Figure~\ref{fig:jepa_training_strategy} compares H-JEPA training settings
under the same forecasting architecture on both datasets. M4 first pretrains
its encoder with H-JEPA and then trains the complete model using only forecast
supervision. On SEVIR, M4 raises CSI-last from 0.1374 to 0.1421, while
CSI$_{219}$ changes from 0.0342 to 0.0345. Joint training further improves
these metrics by 5.1\% and 9.0\%. The same trend holds on MeteoNet. M4 raises
CSI$_{32}$ from 0.1286 to 0.1338 and CSI-last from 0.1402 to 0.1441; joint
training reaches 0.1409 and 0.1734. These results support using H-JEPA as a
joint regularizer rather than only for encoder pretraining. Keeping its
objective active lets the encoder learn from historical prediction and
forecast supervision together, reducing the risk that forecast-only training
weakens the predictive relationships learned from history.
\end{samepage}

\section{Conclusion}
\label{sec:conclusion}

In this paper, we presented \method{} for long-term deterministic precipitation
nowcasting.  The method uses history-masked latent prediction to shape the
shared radar representation, converts it into lead-specific future states, and
renders future frames through motion and source--sink fields.
Experiments on SEVIR and MeteoNet demonstrate improved overall,
high-intensity, and long-term forecasting performance.  Ablation studies
further support the complementary roles of the three components.  Future work
will extend the framework to probabilistic forecasting.

% The optional fifth page contains only material permitted by the conference.
\noindent\textbf{Compliance with Ethical Standards}\par\smallskip
This study uses public meteorological radar datasets and involves no human or
animal subjects; ethical approval was therefore not required. The authors
declare no conflicts of interest and received no funding.

\ifincludereferences
  \bibliographystyle{IEEEbib}
  \bibliography{egbib}
\fi

\end{document}